# HESTENES' TETRAD AND SPIN CONNECTIONS

Frank Reifler and Randall Morris

Lockheed Martin Corporation MS2 (137-205)

199 Borton Landing Road

Moorestown, New Jersey 08057

**ABSTRACT**

Defining a spin connection is necessary for formulating Dirac's bispinor equation in a curved space-time. Hestenes has shown that a bispinor field is equivalent to an orthonormal tetrad of vector fields together with a complex scalar field. In this paper, we show that using Hestenes' tetrad for the spin connection in a Riemannian space-time leads to a Yang-Mills formulation of the Dirac Lagrangian in which the bispinor field $\Psi$ is mapped to a set of $SL(2,R) \times U(1)$ gauge potentials $F_\alpha^K$ and a complex scalar field $\rho$. This result was previously proved for a Minkowski space-time using Fierz identities. As an application we derive several different non-Riemannian spin connections found in the literature directly from an arbitrary linear connection acting on the tensor fields $(F_\alpha^K, \rho)$. We also derive spin connections for which Dirac's bispinor equation is form invariant. Previous work has not considered form invariance of the Dirac equation as a criterion for defining a general spin connection.

## I. INTRODUCTION

Defining a spin connection to replace the partial derivatives in Dirac's bispinor equation in a Minkowski space-time, is necessary for the formulation of Dirac's bispinor equation in a curved space-time. All the spin connections acting on bispinors found in the literature first introduce a local orthonormal tetrad field on the space-time manifold and then require that the Dirac Lagrangian be invariant under local change of tetrad [1] – [8]. Invariance of the Lagrangian by itself does not uniquely define the spin connection [7]. In this paper we determine the spin connections for which the Dirac equation is form invariant. Form invariance means that the Dirac equation can be expressed solely with the spin connection, with no additional terms involving the tetrad, torsion, or non-metricity tensors [7]. We show that such spin connections exist for all linear connections. Form invariance of Dirac's bispinor equation has not generally been considered as a criterion for defining a spin connection [1] – [8].

Using geometric algebra, Hestenes showed in 1967 that a bispinor field on a Minkowski space-time is equivalent to an orthonormal tetrad of vector fields together with a complex scalar field, and that fermion plane waves can be represented as rotational modes of the tetrad [9]. More recently, the Dirac and Einstein equations were unified in a tetrad formulation of a Kaluza-Klein model which gives precisely the usual Dirac-Einstein Lagrangian [10] – [12]. In this model, the self-adjoint (symmetric) modes of the tetrad describe gravity, whereas, as in Hestenes' work, the isometric (rotational) modes of the tetrad together with a scalar field describe fermions. An analogy can be made between the tetrad modes and the elastic and rigid modes of a deformable body [11]. For a deformable body, the elastic modes are self-adjoint and the rigid modes are isometric with respect to the Euclidean metric on $R^3$. This analogy extends into the quantum realm since rigid modes satisfying Euler's equation can be Fermi quantized [13], [14].

To define bispinors, even in a Minkowski space-time, a reference tetrad or its equivalent (e.g., a normal coordinate basis) must first be defined at each point of the space-time [2] – [5], [15] – [17]. Note that the use of such reference tetrads has a long history, dating back to Weyl's 1929 paper [6]. To show that the Dirac bispinor Lagrangian depends only on a tetrad and a scalar field, requires an appropriate choice of reference tetrad. The appropriate choice is provided by Hestenes' orthonormal tetrad of vector fields, denoted as $e_a^\alpha$, where $\alpha = 0, 1, 2, 3$ is a space-time index and $a = 0, 1, 2, 3$ is a tetrad index [9]. Relative to this special reference tetrad, a bispinor field $\Psi$ is "at rest" at each space-time point and has components given as follows (see Section 2):

$$\Psi = \begin{bmatrix} 0 \\ \text{Re}[\sqrt{s}] \\ 0 \\ -i\,\text{Im}[\sqrt{s}] \end{bmatrix} \tag{1.1}$$

where s is a complex scalar field defined in Section 2 by formula (2.4). Note that Hestenes' tetrad $e_a^\alpha$ and the complex scalar field $\sqrt{s}$ are smoothly defined locally in open regions about each space-time point where s is nonvanishing. In each of these open regions, since the Dirac bispinor Lagrangian in a Riemannian space-time depends only on the reference tetrad and (quadratically) on the bispinor field $\Psi$, we show in Section 2 using formula (1.1) that the Dirac bispinor Lagrangian can be expressed entirely in terms of the tensor fields $e_a^\alpha$ and s, once Hestenes' tetrad has been chosen as the reference.

Whenever $\Psi$ vanishes, both s and its first partial derivatives vanish. Setting s and its first partial derivatives to zero in the tensor form of Dirac's bispinor equation shows that $e_a^\alpha$ can be chosen arbitrarily at all space-time points where $\Psi$ vanishes. Thus, all aspects of Dirac's bispinor equation are faithfully reflected in the tensor equations (see Section 2). Since the tetrad $e_a^\alpha$ is unconstrained by the Dirac equation when $\Psi$ vanishes, a gravitational field exists even if the fermion field vanishes. We showed in previous work that the gravitational field $g_{\alpha\beta}$ and the bispinor field $\Psi$ (which together have 10 + 8 = 18 real components), are represented accurately by Hestenes' tensor fields $e_a^\alpha$ and s (which also have 16 + 2 = 18 real components) [10] – [12].

Hestenes' tetrad has been of interest for other applications. Zhelnorovich used Hestenes' tetrad together with the bispinor field at rest as in formula (1.1) to derive spatially flat Bianchi type I solutions of the Einstein-Dirac equations [18], [19]. Hestenes' tetrad in this application has the advantages of reducing the number of unknowns by six and of not requiring special symmetry directions for choosing the tetrad, which considerably simplifies the Einstein-Dirac equations for non-diagonal metrics and makes it possible to obtain new exact solutions [18], [19].

It might seem that Hestenes' tensor fields do not lead to a well posed initial value problem when isolated parts of a bispinor field, with disjoint (closed) supports in a Minkowski space-time, are rotated 360 degrees relative to one another [13], [20]. However, no such isolation is possible because physical bispinor fields with energy bounded from below have supports filling all of space-time, and thereby the tensor fields determine a physical bispinor field uniquely, up to a single unobservable sign [21] – [24].

The Kaluza-Klein tetrad model is based on a constrained Yang-Mills formulation of the Dirac Theory [10] – [12]. In this formulation Hestenes' tensor fields $e_a^\alpha$ and s are mapped bijectively onto a set of $SL(2,R) \times U(1)$ gauge potentials $F_\alpha^K$ and a complex scalar field $\rho$. Thus we have the composite map $\Psi \to (e_a^\alpha, s) \to (F_\alpha^K, \rho)$ (see Section 2). The fact that $e_a^\alpha$ is an orthonormal tetrad of vector fields imposes an orthogonal constraint on the gauge potentials $F_\alpha^K$ given by

$$F_\alpha^K F_{K\beta} = |\rho|^2 g_{\alpha\beta} \qquad (1.2)$$

where $g_{\alpha\beta}$ denotes the space-time metric. The gauge index K = 0, 1, 2, 3 is lowered and raised using a gauge metric $g_{JK}$ and its inverse $g^{JK}$ (see Section 2). Repeated indices are summed. We show in Section 2 that via the map $\Psi \to (F_\alpha^K, \rho)$ the Dirac bispinor Lagrangian equals the following Yang-Mills Lagrangian for the gauge potentials $F_\alpha^K$ and complex scalar field $\rho$ satisfying the orthogonal constraint (1.2), in the limit of an infinitely large coupling constant which we denote as g:

$$L_g = \frac{1}{4g} F^K_{\alpha\beta} F_K^{\alpha\beta} + \frac{1}{g_0} \overline{D_\alpha(\rho+\mu)} D^\alpha(\rho+\mu) \qquad (1.3)$$

where $F^K_{\alpha\beta}$ is the Yang-Mills field tensor with self coupling g, and $D_\alpha$ is the Yang-Mills covariant derivative acting on the scalar field $\rho$ and mass parameter $\mu$. Moreover, $\rho$ and $\mu$ are coupled to the U(1) gauge potential $F^3_\alpha$ with coupling constant $g_0 = (3/2)g$, and $\mu = 2m_0/g_0$ where $m_0$ is the fermion mass (see Section 2). In the limit that g becomes infinitely large, $L_g$ equals Dirac's bispinor Lagrangian.

In Section 3 we reverse our steps by substituting a general linear connection for the Riemannian connection in the Yang-mills Lagrangian (1.3), and thereby derive Dirac's bispinor Lagrangian for space-times with general linear connections. From this Lagrangian we obtain spin connections, for space-times with general linear connections, that satisfy the following two conditions:

1) The tensor and bispinor Lagrangians are equal.
2) The bispinor Dirac equation is form invariant.

We show that such spin connections exist for all linear connections. While spin connections $\nabla_a$ satisfying conditions (1) and (2) are not unique, we prove that the Dirac operators $D = \gamma^a \nabla_a$ formed by them are unique (where $\gamma^a$ are the Dirac matrices [25]). Finally, in Section 3 we relate the spin connections derived from the tensor theory to several different spin connections discussed in the literature that do not satisfy conditions (1) and (2). These spin connections in the literature give Dirac operators different from the unique Dirac operators derived from the tensor theory.

## 2. HESTENES' TETRAD AND THE TENSOR FORM OF THE DIRAC LAGRANGIAN

Even in a Minkowski space-time, bispinors require a reference tetrad for their definition. Other authors have noted that because the Dirac gamma matrices are regarded as constant matrices, the Dirac equation, as described in most textbooks, is not covariant even under Lorentz transformations in the usual sense [26]. Covariant tensor forms of the Dirac bispinor Lagrangian were derived by Zhelnorovich [27] and by Takahashi [28], [29], using trace formulas of the Dirac matrices known as Fierz identities [30], [31]. A simpler derivation using trace formulas of the Pauli matrices was presented as Appendix A and B of reference [32]. In this section we will give a straightforward derivation of the tensor form of the Dirac Lagrangian by using Hestenes' tetrad [9] as the reference tetrad for the spin connection in a Riemannian space-time. For those familiar with spin connections [4], this derivation will be the most direct. As in previous work, we show that the Dirac bispinor Lagrangian equals a constrained Yang-Mills Lagrangian for the gauge group $SL(2,R) \times U(1)$ in the limit of an infinitely large coupling constant. Both the constraint and the limit are explicated in the Kaluza-Klein model [10] – [12].

At each point of a four dimensional Riemannian space-time, bispinors are defined relative to a reference tetrad of orthonormal vectors [2] – [5]. Usually in a Minkowski space-time the reference tetrad consists of coordinate vector fields associated with Cartesian coordinates, but this special choice of reference tetrad is not necessary. A general reference tetrad will be denoted by $e_a$ where a = 0, 1, 2, 3 is a tetrad index. We can express the tetrad $e_a$ as $e_a = e_a^\alpha \partial_\alpha$ where $\partial_\alpha$ for $\alpha$ = 0, 1, 2, 3 denote the partial derivatives with respect to local space-time coordinates $x^\alpha$, and $e_a^\alpha$ denote the tensor components of $e_a$. Tensor indices $\alpha, \beta, \gamma, \delta$ are lowered and raised using the space-time metric, denoted as $g_{\alpha\beta}$, and its inverse $g^{\alpha\beta}$. Tetrad indices a, b, c, d are lowered and raised using a Minkowski metric $g_{ab}$ (with diagonal elements {1, −1, −1, −1} and zeros off the diagonal), and its inverse $g^{ab}$. Repeated tensor and tetrad indices will be summed from 0 to 3.

Using a reference tetrad $e_a$, the spin connection $\nabla_a$ acting on a bispinor field $\Psi$ in a Riemannian space-time is given by [4]:

$$\nabla_a = e_a^\alpha \partial_\alpha - \frac{i}{4} e_a^\alpha e_b^\beta (\nabla_\alpha e_{\beta c}) \sigma^{bc} \qquad (2.1)$$

where

$$\sigma^{bc} = \frac{i}{2}(\gamma^b \gamma^c - \gamma^c \gamma^b) \qquad (2.2)$$

and where $\nabla_\alpha$ denotes the Riemannian connection acting on the vector fields $e_a$, and $\gamma^a$ are (constant) Dirac matrices. (Definitions and sign conventions for the Dirac matrices in this paper will be as in Bjorken and Drell [25].)

Dirac's bispinor Lagrangian in a Riemannian space-time is given by [2], [5]:

$$L_D = \text{Re}[i\overline{\Psi}\gamma^a \nabla_a \Psi - m_0 s] \qquad (2.3)$$

where $m_0$ denotes the fermion mass and the complex scalar field s is defined by

$$\text{Re}[s] = \overline{\Psi}\Psi$$
$$\text{Im}[s] = i\overline{\Psi}\gamma^5\Psi \qquad (2.4)$$

where (using bispinor notation) $\overline{\Psi} = \Psi^+ \gamma^0$, where $\Psi^+$ denotes the transpose conjugate of $\Psi$, and $\gamma^5 = i\gamma^0 \gamma^1 \gamma^2 \gamma^3$ is the fifth Dirac matrix [25]. Formula (2.3) generalizes the usual Dirac bispinor Lagrangian for a Minkowski space-time which uses the coordinate reference tetrad $e_a = \delta_a^\alpha \partial_\alpha$, where $\delta_a^\alpha$ equals one if $a = \alpha$ and zero otherwise. In Theorem 1, a different choice of reference tetrad $e_a$ (Hestenes' tetrad) will lead to the tensor form of the Dirac Lagrangian.

Except for the mass term, Dirac's bispinor Lagrangian (2.3) is invariant under $SL(2,R) \times U(1)$ gauge transformations acting on the bispinor field $\Psi$, with infinitesimal generators $\tau_K$ for K = 0, 1, 2, 3 defined by [32], [33]:

$$\tau_0 \Psi = -i\Psi, \qquad \tau_1 \Psi = i\Psi^C$$
$$\tau_2 \Psi = \Psi^C, \qquad \tau_3 \Psi = i\gamma^5 \Psi$$
(2.5)

where $\Psi^C$ denotes the charge conjugate of $\Psi$ (using bispinor notation [25]). Note that the action of $SL(2,R) \times U(1)$ on $\Psi$ is real linear, whereas usually only complex linear gauge transformations of bispinors are considered. The infinitesimal gauge generators $\tau_0, \tau_1, \tau_2$ generate SL(2,R), and $\tau_3$ generates U(1).

The $SL(2,R) \times U(1)$ gauge transformations generated by $\tau_K$ commute with Lorentz transformations [25]. From formula (2.5) the commutation relations of the gauge generators $\tau_K$ are given by

$$[\tau_0, \tau_1] = 2\tau_2, \qquad [\tau_0, \tau_2] = -2\tau_1$$
$$[\tau_1, \tau_2] = -2\tau_0$$
(2.6)

and $\tau_3$ commutes with all the $\tau_K$. Formula (2.6) can be expressed as

$$[\tau_J, \tau_K] = 2f_{JK}^L \tau_L \tag{2.7}$$

where $f_{JK}^L$ are the Lie algebra structure constants for the gauge group $SL(2,R) \times U(1)$. Note that from formula (2.6):

$$f_{JKL} = g_{LM} f_{JK}^{M} = -\varepsilon_{JKL3} \tag{2.8}$$

where $g_{LM}$ is the Minkowski metric (with diagonal elements $\{1,-1,-1,-1\}$ and zeros off the diagonal), and $\varepsilon_{JKLM}$ is the permutation tensor (with $\varepsilon_{0123} = -\varepsilon^{0123} = 1$). Gauge indices J, K, L, M are lowered and raised using the gauge metric $g_{JK}$, and its inverse $g^{JK}$. Repeated gauge indices are summed from 0 to 3.

The scalar field s in formula (2.4) is invariant under SL(2,R) gauge transformations, and transforms as a complex U(1) scalar under the U(1) gauge transformations (i.e., chiral gauge transformations [34]). To make the Lagrangian (2.3) invariant for all SL(2,R)×U(1) gauge transformations, it suffices that $m_0$ transform like $\bar{s}$ (the complex conjugate of s). Since $m_0$ appears in the Lagrangian (2.3) without derivatives, the assumption that $m_0$ transform like $\bar{s}$ under U(1) chiral gauge transformations, has no effect on the Dirac equation.

From the Dirac bispinor Lagrangian (2.3) we can derive the following SL(2,R)×U(1) Noether currents $j^K = j_a^K e^a$ with tetrad components:

$$j_a^K = \text{Re}[i\bar{\Psi}\gamma_a \tau^K \Psi] \tag{2.9}$$

Note that, $j^0$, $j^1$, and $j^2$ are SL(2,R) Noether currents and $j^3$ is the U(1) Noether current. In particular $j^0$ is the electromagnetic current and $j^3$ is the chiral current; i.e.,

$$\begin{aligned} j_a^0 &= \bar{\Psi}\gamma_a \Psi \\ j_a^3 &= \bar{\Psi}\gamma_a \gamma^5 \Psi \end{aligned} \tag{2.10}$$

whereas [28], [29],

$$j_a^1 = \text{Re}[\overline{\Psi}\gamma_a\Psi^C]$$
$$j_a^2 = \text{Im}[\overline{\Psi}\gamma_a\Psi^C]$$
(2.11)

where $j_a^K$ denote the tetrad components of $j^K = j_a^K e^a$. The real Noether currents $j^K$ and complex scalar field s satisfy an orthogonal constraint known as a Fierz identity [28], [29]:

$$j_a^K j_{Kb} = |s|^2 g_{ab}$$
$$j_a^J j^{Ka} = |s|^2 g^{JK}$$
(2.12)

A derivation of the tensor form of Dirac's bispinor Lagrangian (2.3) follows from the map $\Psi \to (j_a^K, s)$. Apart from the singular set where s vanishes, we can make a special choice of orthonormal reference tetrad as follows:

$$e_a = |s|^{-1} \delta_a^K j_K$$
(2.13)

The following lemma shows that relative to this special reference tetrad, which is Hestenes' tetrad [9], the bispinor field $\Psi$ at each point in the space-time is "at rest", and $\Psi$ becomes locally a function of a complex scalar field $\sigma$, which has s as its square.

**LEMMA 1:**

Relative to Hestenes' tetrad (2.13), at each space-time point where Hestenes' tetrad is defined, every bispinor field $\Psi$ has the form:

$$\Psi = \begin{bmatrix} 0 \\ \mathrm{Re}[\sigma] \\ 0 \\ -i\,\mathrm{Im}[\sigma] \end{bmatrix} \qquad (2.14)$$

where $\sigma$ is a locally defined complex scalar field, which has s as its square.

**PROOF:**

Given $j^K$ and s, we will solve for $\Psi$. Substituting $j^K$ defined by formula (2.9) into formula (2.13) for Hestenes' tetrad, gives

$$\mathrm{Re}[i\overline{\Psi}\gamma_a \tau^K \Psi] = |s|\delta_a^K \qquad (2.15)$$

It is then straightforward to verify that all solutions of equations (2.4) and (2.15) are of the form (2.14) with the complex scalar $\sigma$ having s as its square.  Q.E.D.

Note that choosing Hestenes' tetrad as the reference tetrad reduces the bispinor field $\Psi$ to locally depend only on a scalar field $\sigma$, at all points where Hestenes' tetrad is defined. Substitution of formula (2.14) for $\Psi$ into formula (2.3), expresses the Dirac bispinor Lagrangian in terms of Hestenes' tensor fields $(e_a, \sigma)$. Further examination of formulas (2.1), (2.3), and (2.14) shows that the Dirac Lagrangian can be expressed solely with the tensor fields $(j^K, s)$. This result, proved below in Theorem 1, was first derived by Takahashi using Fierz identities [28], [29].

To show that the tensor form of Dirac's bispinor Lagrangian (2.3) is a constrained Yang-Mills Lagrangian in the limit of an infinitely large coupling constant, we map $SL(2,R) \times U(1)$ gauge potentials $F_\alpha^K$ and a complex scalar field $\rho$ into $(j^K, s)$ by setting:

$$j_\alpha^K = 4|\rho|^2 F_\alpha^K \qquad (2.16)$$

$$s = 4|\rho|^2 \bar{\rho}$$

where $j_\alpha^K = j_a^K e_\alpha^a$ are the tensor components of $j^K$. From formulas (2.12) and (2.16), since the reference tetrad $e_a$ is orthonormal, the gauge potentials $F_\alpha^K$ satisfy an orthogonal constraint, which can be expressed in two equivalent ways:

$$F_\alpha^K F_{K\beta} = |\rho|^2 g_{\alpha\beta} \qquad (2.17)$$

$$F_\alpha^J F^{K\alpha} = |\rho|^2 g^{JK}$$

Consider the following Yang-Mills Lagrangian for the gauge potentials $F_\alpha^K$ and the complex scalar field $\rho$:

$$L_g = \frac{1}{4g} F_{\alpha\beta}^K F_K^{\alpha\beta} + \frac{1}{g_0} \overline{D_\alpha(\rho+\mu)} D^\alpha(\rho+\mu) \qquad (2.18)$$

where, because of the symmetry of the Riemannian connection, the Yang-Mills field tensor $F_{\alpha\beta}^K$ is given by

$$F_{\alpha\beta}^K = \nabla_\alpha F_\beta^K - \nabla_\beta F_\alpha^K + g f_{MN}^K F_\alpha^M F_\beta^N \qquad (2.19)$$

and where the Yang-Mills coupling constant g is a self-coupling of the gauge potentials $F_\alpha^K$. Furthermore, in the Lagrangian (2.18), the complex scalar $\mu$ satisfies:

$$\mu = \frac{2m_0}{g_0}, \qquad \partial_\alpha \mu = 0 \qquad (2.20)$$

where $m_0$ is the fermion mass, and $g_0 = (3/2)g$. As previously stated for Dirac's bispinor Lagrangian (2.3) both the complex scalar field s and the fermion mass $m_0$ transform as U(1) scalars. The same is true for $\rho$ and $\mu$ by formulas (2.16) and (2.20). Hence the covariant derivative $D_\alpha$ acts on $\rho + \mu$ as follows:

$$D_\alpha(\rho+\mu) = \partial_\alpha \rho - ig_0 F_\alpha^3 (\rho+\mu) \qquad (2.21)$$

That is, $g_0 = (3/2)g$ is the Yang-Mills constant which couples the U(1) scalars $\rho$ and $\mu$ to the U(1) gauge potential $F_\alpha^3$. (By formula (2.8), the Lie algebra structure constants $f_{JK}^L$ vanish if any gauge index J, K, L equals 3, so that $g_0$ can be different than g.) Note that the complex scalar field $\mu$ acts as a Higgs field for generating the fermion mass $m_0$. We could consider subtracting a large quartic potential containing $\mu$ from the Yang-mills Lagrangian (2.18) of the form [34]:

$$V_g(\mu) = g_0^6 \left( |\mu|^2 - \frac{4|m_0|^2}{g_0^2} \right)^2 \qquad (2.22)$$

whereby formulas (2.20) and (2.21) become effective formulas for large $g_0$.

**THEOREM 1:**

At every space-time point where Hestenes' tetrad is defined, Dirac's bispinor Lagrangian (2.3) equals the Yang-Mills Lagrangian (2.18) in the limit of a large coupling constant. That is,

$$L_D = \lim_{g \to \infty} L_g \qquad (2.23)$$

**PROOF:**

From formulas (2.1), (2.2), (2.3), (2.10), and using the following identity for Dirac matrices [5], [35]:

$$\gamma^a \gamma^b \gamma^c = g^{ab}\gamma^c - g^{ac}\gamma^b + g^{bc}\gamma^a + i\varepsilon^{abcd}\gamma^5\gamma_d \qquad (2.24)$$

we can express Dirac's bispinor Lagrangian in a Riemannian space-time as a sum of three terms:

$$L_D = \text{Re}[i\overline{\Psi}\gamma^a e_a^\alpha \partial_\alpha \Psi - m_0 s] + \frac{1}{4}\varepsilon^{abcd} e_a^\alpha e_b^\beta (\nabla_\alpha e_{\beta c}) j_d^3 \qquad (2.25)$$

We will express each of these terms with the tensor fields $(F_\alpha^K, \rho)$. Substituting formula (2.14) for $\Psi$ in the first term of $L_D$, we get using formula (2.16),

$$\text{Re}[i\overline{\Psi}\gamma^a e_a^\alpha \partial_\alpha \Psi] = \text{Re}[2i\overline{\rho} F_\alpha^3 \partial^\alpha \rho] \qquad (2.26)$$

For the second term of $L_D$, formula (2.16) gives

$$-\text{Re}[m_0 s] = -\text{Re}[4m_0 |\rho|^2 \overline{\rho}] \qquad (2.27)$$

Noting that $j_a^K = |s|\delta_a^K$ by formulas (2.9) and (2.15), and using formulas (2.13) and (2.16), the third term of $L_D$ becomes:

$$\frac{1}{4}\varepsilon^{abcd} e_a^\alpha e_b^\beta (\nabla_\alpha e_{\beta c}) j_d^3 = -(\nabla_\alpha \mathbf{F}_\beta) \bullet \mathbf{F}^\alpha \times \mathbf{F}^\beta \qquad (2.28)$$

where $\mathbf{F}_\alpha = (F_\alpha^0, F_\alpha^1, F_\alpha^2)$. Summing the three terms (2.26), (2.27), and (2.28), formula (2.25) becomes:

$$L_D = \text{Re}[-(\nabla_\alpha \mathbf{F}_\beta) \bullet \mathbf{F}^\alpha \times \mathbf{F}^\beta + 2i\bar{\rho}F_\alpha^3(\nabla^\alpha \rho) - 4m_0|\rho|^2\bar{\rho}] \qquad (2.29)$$

Terms in the Yang-Mills Lagrangian (2.18) which are quartic in the fields $(F_\alpha^K, \rho)$ vanish by virtue of the orthogonal constraint (2.17) and the relation between the coupling constants $g_0 = (3/2)g$. Quadratic terms in the fields $(F_\alpha^K, \rho)$ vanish in the limit (2.23). Thus, the limit (2.23) only contains terms cubic in the fields $(F_\alpha^K, \rho)$. The cubic terms of the Yang-Mills Lagrangian (2.18) are given by

$$L_g^{(3)} = \text{Re}[f_{JKL}(\nabla_\alpha F_\beta^J)F^{K\alpha}F^{L\beta} + 2i\bar{\rho}F_\alpha^3(\nabla^\alpha \rho) + 4m_0 F_\alpha^3 F^{3\alpha}\bar{\rho}] \qquad (2.30)$$

which equals $L_D$ given in formula (2.29), after applying the orthogonal constraint (2.17) to obtain

$$F_\alpha^3 F^{3\alpha} = -|\rho|^2 \qquad (2.31)$$

and using formula (2.8) to replace the triple vector product with the Lie algebra structure constants $f_{JKL}$. Q.E.D.

Since Theorem 1 shows that the Dirac bispinor Lagrangian (2.3) and its tensor form (2.29) are equal at all space-time points where Hestenes' tetrad is defined, we will briefly discuss the physical interpretation of the singularities, where Hestenes' tetrad is not defined. By formula (2.13) Hestenes' tetrad $e_a$ is defined wherever the scalar field s does not vanish. When s vanishes there are two types of singularities. First, if the bispinor field $\Psi$ vanishes, both s and its first partial derivatives vanish by formula (2.4), and the tensor form of the Dirac equation allows $e_a$ to be arbitrary. At such space-time points the tensor fields $F_\alpha^K$ and $\rho$ in the Lagrangian (2.18) vanish. Second, if s vanishes but $\Psi$ does not, then the nonvanishing fermion particle current lies on the light cone [9]. For physical solutions representing massive fermions, these singularities must form an exceptional (nowhere dense) set. Thus singularities in the tensor fields $F_\alpha^K$ and $\rho$ can only occur in the complement of an open dense subset of the space-time. Consequently, putative differences between the bispinor field $\Psi$ and the tensor fields $F_\alpha^K$ and $\rho$ cannot be observed in experiments.

Field observables can be derived from Noether's theorem in a Minkowski space-time [36]. Besides the Noether currents $j_\alpha^K$ given in formula (2.16) which are derived from $SL(2,R) \times U(1)$ gauge symmetry, from Minkowski space-time symmetries we obtain the energy-momentum tensor $T_{\alpha\beta}$ and the spin polarization tensor $S_{\alpha\beta\gamma}$, which we will use in Section 3 to derive spin connections acting on bispinor fields. There are three methods for deriving such formulas. First, we can apply Noether's theorem to the Dirac bispinor Lagrangian (2.3) and then use Fierz identities to derive the tensor formulas [32], [33]. Second, we can apply Noether's theorem directly to the tensor form of Dirac's Lagrangian (2.29). Third, we can apply Noether's theorem to the Yang-Mills Lagrangian (2.18) and take the limit as the coupling constant g becomes infinite. For example, for the spin polarization tensor $S_{\alpha\beta\gamma}$ in a Minkowski space-time we have:

$$S_{\alpha\beta\gamma} = -\frac{1}{2}\text{Re}[\overline{\Psi}\gamma_\alpha\sigma_{\beta\gamma}\Psi] = -\frac{1}{2}\varepsilon_{\alpha\beta\gamma\delta}\overline{\Psi}\gamma^\delta\gamma^5\Psi$$
$$= -\frac{1}{2}\varepsilon_{\alpha\beta\gamma\delta}j^{3\delta} = 2\,\mathbf{F}_\alpha \bullet \mathbf{F}_\beta \times \mathbf{F}_\gamma \qquad (2.32)$$

where $\gamma_\alpha = \delta_\alpha^a \gamma_a$ and $\sigma_{\alpha\beta} = (i/2)(\gamma_\alpha\gamma_\beta - \gamma_\beta\gamma_\alpha)$, and where $\varepsilon_{\alpha\beta\gamma\delta}$ denotes the permutation tensor. The expression after the first equals sign in formula (2.32), giving the spin polarization tensor $S_{\alpha\beta\gamma}$ in terms of the generators of Lorentz transformations $-(i/2)\sigma_{\beta\gamma}$, is derived from the bispinor Lagrangian (2.3) in the usual way [37]. The expression after the second equals sign comes from the identity for the gamma matrices (2.24). The expression after the third equals sign uses the definition of the Noether current $j_\alpha^3$. The last expression in formula (2.32) is derived from formula (2.16) and the orthogonal constraint (2.17).

On the other hand, formula (2.32) can be obtained directly from the spin polarization tensor of the Yang-Mills Lagrangian (2.18) as follows:

$$S_{\alpha\beta\gamma} = \lim_{g\to\infty} -\frac{1}{g} \text{Re}[F_{\alpha\beta}^K F_{K\gamma} - F_{\alpha\gamma}^K F_{K\beta}]$$

$$= 2\, \mathbf{F}_\alpha \bullet \mathbf{F}_\beta \times \mathbf{F}_\gamma \qquad (2.33)$$

The first equation of formula (2.33) expresses the spin polarization tensor for the Yang-Mills Lagrangain (2.18) by the usual formula [36]. The second equation uses formula (2.19) to obtain the limit for an infinitely large coupling constant g. Note that formula (2.33) also provides a definition of the spin polarization tensor $S_{\alpha\beta\gamma}$ in a Riemannian space-time.

In a similar manner, we get from Lagrangians (2.3) and (2.18), the energy-momentum tensor $T_{\alpha\beta}$ in a Minkowski space-time as follows:

$$T_{\alpha\beta} = \text{Re}[i\overline{\Psi}\gamma_\alpha \partial_\beta \Psi]$$

$$= \text{Re}[-(\partial_\beta \mathbf{F}_\gamma) \bullet \mathbf{F}_\alpha \times \mathbf{F}^\gamma + 2iF_\alpha^3 \overline{\rho}(\partial_\beta \rho)] \qquad (2.34)$$

This formula agrees with the formula derived by Takahashi [13], [28], [29]. In a Riemannian space-time we define a (non-symmetric) energy-momentum tensor $T_{\alpha\beta}$ by the following formula whose proof is similar to the proof of formula (2.29):

$$T_{\alpha\beta} = e_\alpha^a e_\beta^b\, \text{Re}[i\overline{\Psi}\gamma_a \nabla_b \Psi]$$

$$= \text{Re}[-(\nabla_\beta \mathbf{F}_\gamma) \bullet \mathbf{F}_\alpha \times \mathbf{F}^\gamma + 2iF_\alpha^3 \bar{\rho}(\nabla_\beta \rho)\,] \tag{2.35}$$

where $\nabla_a$ is the spin connection (2.1) and (henceforth in this paper) $e_a^\alpha$ denotes an arbitrary tetrad of orthonormal vector fields.

Takahashi's formula (2.34) and formula (2.32) can be generalized to a Riemannian space-time as follows:

$$\text{Re}[i\bar{\Psi}\gamma_a \partial_\beta \Psi] = \text{Re}[-(\partial_\beta \mathbf{F}_c) \bullet \mathbf{F}_a \times \mathbf{F}^c + 2iF_a^3 \bar{\rho}(\partial_\beta \rho)\,] \tag{2.36}$$

$$\text{Re}[\bar{\Psi}\gamma_a \sigma_{bc} \Psi] = -4\,\mathbf{F}_a \bullet \mathbf{F}_b \times \mathbf{F}_c \tag{2.37}$$

where $F_a^K = F_\alpha^K e_a^\alpha$ and $\mathbf{F}_a = (F_a^0, F_a^1, F_a^2)$. Note that since $\gamma_a$ are constant Dirac matrices, the left hand sides of formulas (2.36) and (2.37) depend only on $\Psi$ and not on $e_a^\alpha$. The same is true of the right hand sides, since like $j_a^K$ and s in formulas (2.4) and (2.9), $F_a^K$ and $\rho$ depend only on $\Psi$. Thus, formula (2.36) is simply a restatement of Takahashi's formula (2.34) with different notation. Using formulas (2.1) and (2.37) and the fact that

$$\partial_\beta \mathbf{F}_c = (\nabla_\beta \mathbf{F}_\gamma)e_c^\gamma + \mathbf{F}_\gamma (\nabla_\beta e_c^\gamma) \tag{2.38}$$

it is straightforward to verify that formulas (2.35) and (2.36) are equivalent.

## 3. SPIN CONNECTIONS FOR AN ARBITRARY LINEAR CONNECTION

In Section 2 we derived the following formula for the tensor form of the Dirac Lagrangian using Hestenes' tetrad in the spin connection for a Riemannian space-time:

$$L_D = \text{Re}[-(\nabla_\alpha \mathbf{F}_\beta) \bullet \mathbf{F}^\alpha \times \mathbf{F}^\beta + 2i\bar{\rho}F_\alpha^3(\nabla^\alpha \rho) - 4m_0|\rho|^2\bar{\rho}] \qquad (3.1)$$

In this section we will reverse our steps and generalize $\nabla_\alpha$ in this tensor Lagrangian from a Riemannian connection to a general linear connection, and from this substitution obtain general spin connections acting on bispinors. In Theorem 2 we describe spin connections for which Dirac's bispinor equation is form invariant. We will then discuss several different spin connections found in the literature for which Dirac's bispinor equation is not form invariant.

We can express the Lagrangian (3.1) using components $F_a^K = F_\alpha^K e_a^\alpha$ of the gauge potentials $F_\alpha^K$ with respect to an arbitrary tetrad of orthonormal vector fields $e_a^\alpha$. The tetrad of orthonormal vector fields $e_a^\alpha$ satisfies:

$$g_{\alpha\beta} e_a^\alpha e_b^\beta = g_{ab}$$
$$g^{ab} e_a^\alpha e_b^\beta = g^{\alpha\beta} \qquad (3.2)$$

where as in Section 2, tetrad indices are denoted by a, b, c, d = 0, 1, 2, 3 and general coordinate indices are denoted by $\alpha$, $\beta$, $\gamma$, $\delta$ = 0, 1, 2, 3 and where $g_{\alpha\beta}$ is a general space-time metric with inverse $g^{\alpha\beta}$, and

$$g_{ab} = g^{ab} = \begin{bmatrix} 1 & 0 & 0 & 0 \\ 0 & -1 & 0 & 0 \\ 0 & 0 & -1 & 0 \\ 0 & 0 & 0 & -1 \end{bmatrix} \quad (3.3)$$

In the derivation that follows, we will be careful to distinguish the covariant and contravariant coordinate indices. Since for a non-Riemannian connection, $\nabla_\alpha g_{\beta\gamma}$ does not generally vanish, notation must distinguish between a tetrad of orthonormal vector fields $e_a^\alpha$, and its dual tetrad of orthonormal one-forms $\varepsilon_\alpha^a$. However, since $g_{ab}$ is a constant metric, we will freely raise and lower tetrad indices (e.g., $e^{\alpha a} = g^{ab} e_b^\alpha$). We have

$$\varepsilon_\alpha^a = g^{ab} g_{\alpha\beta} e_b^\beta$$
$$e_a^\alpha = g^{\alpha\beta} g_{ab} \varepsilon_\beta^b \quad (3.4)$$

Note that $e_a^\alpha$ are components of the vector fields $e_a = e_a^\alpha \partial_\alpha$ with respect to coordinate vector fields $\partial_\alpha$ on the space-time, whereas $\varepsilon_\alpha^a$ are components of the one-forms $\varepsilon^a = \varepsilon_\alpha^a dx^\alpha$ (dual to $e_a$) with respect to coordinate one-forms $dx^\alpha$. Thus,

$$\varepsilon_\alpha^a e_a^\beta = \delta_\alpha^\beta$$
$$\varepsilon_\alpha^a e_b^\alpha = \delta_b^a \quad (3.5)$$

where $\delta_\alpha^\beta$ (respectively $\delta_b^a$) equals one if $\alpha = \beta$ (respectively a = b) and equals zero otherwise. We have

$$F_a^K = F_\alpha^K \, e_a^\alpha$$

$$F_\alpha^K = F_a^K \, \varepsilon_\alpha^a$$

(3.6)

From formulas (3.5) and (3.6), the orthogonal constraint (2.17) becomes:

$$F_a^K \, F_{Kb} = |\rho|^2 \, g_{ab}$$

$$F_a^J \, F^{Ka} = |\rho|^2 \, g^{JK}$$

(3.7)

Note that the components $F_a^K$ transform covariantly under Lorentz and $SL(2,R) \times U(1)$ gauge transformations, but transform as scalars under coordinate transformations.

In a curved space-time, we replace partial derivatives $\partial_\alpha$ with covariant derivatives $\nabla_\alpha$ given by

$$\nabla_\alpha F_\beta^K = \partial_\alpha F_\beta^K - \Gamma_{\alpha\beta}^\gamma \, F_\gamma^K$$

$$\nabla_\alpha \rho = \partial_\alpha \rho$$

(3.8)

where $\Gamma_{\alpha\beta}^\gamma$ are the connection coefficients. We first consider torsion free connections. Since $\Gamma_{\alpha\beta}^\gamma = \Gamma_{\beta\alpha}^\gamma$ for torsion free connections, the Lagrangian (3.1) is unaffected when $\partial_\alpha$ is used instead of $\nabla_\alpha$. The first term of the Lagrangian (3.1) becomes, using formulas (3.2) through (3.8):

$$(\nabla_\alpha \mathbf{F}_\beta) \bullet \mathbf{F}^\alpha \times \mathbf{F}^\beta = (\partial_\alpha \mathbf{F}_\beta) \bullet \mathbf{F}^\alpha \times \mathbf{F}^\beta$$

$$= \partial_\alpha (\mathbf{F}_a \varepsilon_\beta^a) \bullet \mathbf{F}_b e^{\alpha b} \times \mathbf{F}_c e^{\beta c}$$

$$= (\partial_\alpha \mathbf{F}^c) \bullet \mathbf{F}_b \times \mathbf{F}_c e^{\alpha b} + \mathbf{F}_a \bullet \mathbf{F}_b \times \mathbf{F}_c (\partial_\alpha \varepsilon_\beta^a) e^{\alpha b} e^{\beta c}$$

$$= (\partial_\alpha \mathbf{F}^c) \bullet \mathbf{F}_b \times \mathbf{F}_c e^{\alpha b} + \frac{1}{2} \omega_{abc} S^{abc} \tag{3.9}$$

where we denote $\mathbf{F}_a = (F_a^0, F_a^1, F_a^2)$, and similar to formula (2.32) we define:

$$S^{abc} = 2 \mathbf{F}^a \bullet \mathbf{F}^b \times \mathbf{F}^c \tag{3.10}$$

Then using the antisymmetry of $S^{abc}$, we define $\omega_{abc}$ in formula (3.9) as:

$$\omega_{abc} = e_a^\alpha e_b^\beta (\partial_\alpha \varepsilon_{\beta c}) \tag{3.11}$$

Formula (3.9) gives:

$$(\nabla_\alpha \mathbf{F}_\beta) \bullet \mathbf{F}^\alpha \times \mathbf{F}^\beta = e^{\alpha b} (\partial_\alpha \mathbf{F}^c) \bullet \mathbf{F}_b \times \mathbf{F}_c + \frac{1}{2} \omega_{abc} S^{abc} \tag{3.12}$$

From formulas (2.4), (2.16), and (2.36) we have:

$$\mathrm{Re}[i\overline{\Psi}\gamma_b \partial_\alpha \Psi] = \mathrm{Re}[-(\partial_\alpha \mathbf{F}_c) \bullet \mathbf{F}_b \times \mathbf{F}^c + 2i F_b^3 \overline{\rho}(\partial_\alpha \rho)]$$

$$\mathrm{Re}[\overline{\Psi}\Psi] = \mathrm{Re}[4|\rho|^2 \rho] \tag{3.13}$$

Substituting formulas (3.12) and (3.13) into (3.1), we obtain:

$$L_D = \mathrm{Re}[i\overline{\Psi}\gamma^a e_a^\alpha \partial_\alpha \Psi - m_0 \overline{\Psi}\Psi - \frac{1}{2}\omega_{abc} S^{abc}] \tag{3.14}$$

where we used formulas (2.4) and (2.16) to obtain the mass term. Formulas (3.10) and (2.37) give:

$$S^{abc} = -\frac{1}{2}\text{Re}[\overline{\Psi}\gamma^a\sigma^{bc}\Psi] \qquad (3.15)$$

Thus, formula (3.14) becomes:

$$L_D = \text{Re}[i\overline{\Psi}\gamma^a\nabla_a\Psi - m_0\overline{\Psi}\Psi] \qquad (3.16)$$

where $\nabla_a$ acts on bispinors as:

$$\nabla_a = e_a^\alpha\partial_\alpha - \frac{i}{4}\omega_{abc}\sigma^{bc} \qquad (3.17)$$

Since the spin polarization tensor $S^{abc}$ is antisymmetric in all indices (see formula (3.10)), we can cyclically permute the tetrad indices a, b, c of $\omega_{abc}$ without affecting the Lagrangian (3.14). Hence, $\nabla_a$ in formula (3.17) is not unique, and any linear combination of $\omega_{abc}$, $\omega_{bca}$, and $\omega_{cab}$ whose weights sum to one, can replace $\omega_{abc}$. Variation of the action associated with the Lagrangian (3.16) with respect to the bispinor field $\Psi$ then shows that the Dirac equation is form invariant only if the weights $(1, -1, 1)$ are chosen. Form invariant means that the Dirac equation can be expressed solely with the spin connnection $\nabla_a$ as follows:

$$i\gamma^a\nabla_a\Psi = m_0\Psi \qquad (3.18)$$

with no additional terms involving the tetrad [2], [5].

It is straightfoward to generalize the derivation to connections with torsion, whereby $\omega_{abc}$ in formula (3.17) for $\nabla_a$ becomes:

$$\omega_{abc} = e_a^\alpha e_b^\beta (\partial_\alpha \varepsilon_{\beta c}) - \frac{1}{2} t_{\alpha\beta\gamma} e_{[a}^\alpha e_b^\beta e_{c]}^\gamma \qquad (3.19)$$

where $t_{\alpha\beta\gamma} = \Gamma_{\alpha\beta\gamma} - \Gamma_{\beta\alpha\gamma}$ is the torsion tensor, and the brackets [a, b, c] indicate an antisymmetric average over the tetrad indices a, b, c. Following the previous argument, we can replace $\omega_{abc}$ with any linear combination of $\omega_{abc}$, $\omega_{bca}$, and $\omega_{cab}$ whose weights sum to one, and again the linear combination with weights $(1, -1, 1)$ defines the spin connection for which the Dirac equation is form invariant. Thus,

$$\nabla_a = e_a^\alpha \partial_\alpha - \frac{i}{4}(\Omega_{abc} - \Omega_{bca} + \Omega_{cab} - t_{[abc]}/2)\,\sigma^{bc} \qquad (3.20)$$

where $\Omega_{abc} = e_a^\alpha e_b^\beta (\partial_\alpha \varepsilon_{\beta c})$ and $t_{abc} = t_{\alpha\beta\gamma} e_a^\alpha e_b^\beta e_c^\gamma$. From formula (3.20) we see that this spin connection depends only on the reference tetrad $\varepsilon_\alpha^a$ and the totally antisymmetric part of the torsion tensor $t_{\alpha\beta\gamma}$. When torsion vanishes, this spin connection depends only on the reference tetrad [2] – [5]. Noting that $\sigma^{bc} = -\sigma^{cb}$, formula (3.20) agrees with the spin connections in Hehl and Datta [2] and in Hammond [5] for metric compatible connections with totally antisymmetric torsion.

**THEOREM 2:**

For an arbitrary linear connection $\nabla_\alpha$, the spin connection $\nabla_a$ defined by formula (3.20) satisfies the following two conditions:

1) The bispinor Lagrangian (with $\nabla_a$) equals the tensor Lagrangian (with $\nabla_\alpha$).
2) The bispinor Dirac equation is form invariant.

Furthermore, the Dirac operator $D = \gamma^a \nabla_a$ is unique for all spin connections $\nabla_a$ satisfying conditions (1) and (2).

**PROOF:**

Since the spin connection (3.20) satisfies the two conditions, it remains to prove only the second assertion. Let $\nabla_a$ and $\nabla'_a = \nabla_a + O_a$ be two spin connections satisfying conditions (1) and (2), where the two spin connections differ by an operator $O_a$ acting on bispinors. Let $D = \gamma^a \nabla_a$ and $D' = \gamma^a \nabla'_a$ be the respective Dirac operators, and let $L_D$ and $L'_D$ denote the respective bispinor Lagrangians (3.16). By condition (1), both bispinor Lagrangians equal the tensor Lagrangian (3.1), hence $L'_D = L_D$. We have

$$L'_D - L_D \;=\; \mathrm{Re}[i\overline{\Psi}\gamma^a O_a \Psi] \;=\; 0 \qquad (3.21)$$

From formulas (3.16) and (3.21) the Euler-Lagrange equations give:

$$i\gamma^a \nabla_a \Psi \;=\; m_0 \Psi \;=\; i\gamma^a \nabla'_a \Psi \;=\; i\gamma^a \nabla_a \Psi + i\gamma^a O_a \Psi \qquad (3.22)$$

Thus, $\gamma^a O_a = 0$. Q.E.D.

Note that the operator $O_a$ must be constructed from available tensors, such as torsion and non-metricity tensors. In the Riemannian case, these tensors vanish, so that $O_a = 0$. For example, take the spin connection in Utiyama and Jhangiani [1], [4]:

$$\nabla_a \;=\; e_a^\alpha \partial_\alpha - \frac{i}{4} e_a^\alpha e_b^\beta (\nabla_\alpha \varepsilon_{\beta c}) \sigma^{bc} \qquad (3.23)$$

For a Riemannian connection $\nabla_\alpha$, expressing the Riemannian connection in terms of the tetrad $\varepsilon_\alpha^a$ in formula (3.23), gives the spin connection (3.20) with weights $(1, -1, 1)$ as before. However for a general linear connection, formula (3.23) does not give a form invariant Dirac equation as in formula (3.18). That is, the spin connections (3.20) and (3.23) are not equal for general linear connections.

Note that without changing the Lagrangian (3.16), we can change any spin connection $\nabla_a$ to $\nabla'_a = \nabla_a + v_a$, where $v_a$ is any (real) Lorentz four-vector field. Hurley and Vandyck, studying conformal connections that commute with tensor-spinor maps, consider a class of such spin connections where

$$v_a = -\frac{k}{16} e_a^\alpha g^{\beta\gamma} (\nabla_\alpha g_{\beta\gamma}) \tag{3.24}$$

and where k is a constant [7]. Except in special cases, these spin connections when substituted into the Lagrangian (3.16) do not generally make the Dirac equation form invariant.

De Andrade, Guillen, and Pereira propose a teleparallel spin connection that, though similar to the spin connection (3.20), lacks a torsion term [8]. While this spin connection makes the bispinor Dirac equation form invariant, it is not derived from the teleparallel connection in the tensor theory. Substituting a teleparallel connection $\nabla_\alpha$ into formula (3.1) results in a spin connection containing the torsion term $t_{[abc]}$. Thus, the proposed spin connection is derived from a Riemannian connection in the tensor theory, and not from a teleparallel connection.